\documentclass[twocolumn]{aa}
\usepackage{graphicx}
\usepackage{txfonts}

\def\ms{\,m\,s$^{-1}$}         
\def\kms{\,km\,s$^{-1}$}       

\begin{document}
   \title{Two new ''very hot Jupiters'' \\
   among the OGLE transiting candidates}


   \author{F. Bouchy\inst{1,2}, F. Pont\inst{1}, N.C. Santos\inst{3,2},C. Melo\inst{4}, 
           M. Mayor\inst{2}, D. Queloz\inst{2} \and S. Udry\inst{2}\fnmsep\thanks{Based on observations 
	   collected with the FLAMES+UVES spectrograph at the VLT/UT2 Kueyen telescope 
	   (Paranal Observatory, ESO, Chile)}
          }

   \offprints{\email{Francois.Bouchy@oamp.fr}}

   \institute{Laboratoire d'Astrophysique de Marseille, 
               Traverse du Siphon, 13013 Marseille, France
	  \and
              Observatoire de Gen\`eve, 51 ch. des Maillettes, 1290 Sauverny, Switzerland
         \and
	     Centro de Astronomia e Astro\'{\i}fsica da Universidade de Lisboa,
Observat\'orio Astron\'omico de Lisboa, Tapada da Ajuda, 1349-018 Lisboa, Portugal
         \and
	      ESO, Casilla 19001, Santiago 19, Chile  
            }

   \date{Received ; accepted }

   \abstract{As a result of a radial velocity follow-up of OGLE planetary transit candidates in Carina, 
   we report the discovery of two new transiting planets with very short orbital periods: 
   OGLE-TR-113 with m=1.35 $\pm 0.22 M_{Jup}$, r=$1.08^{+0.07}_{-0.05} R_{Jup}$, P=1.43 day, and 
   OGLE-TR-132 with m=1.01 $\pm 0.31 M_{Jup}$,  r=$1.15^{+0.80}_{-0.13} R_{Jup}$, P=1.69 day. 
   These detections bring to three the number of known "very hot Jupiters" 
   (Jovian exoplanets like OGLE-TR-56 with periods much smaller than 3 days), indicating that the 
   absence of such planets in radial velocity survey does not reflect an absolute limit. 
   
   }

   \authorrunning{F. Bouchy et al.}
   \titlerunning{Two new ''very hot Jupiters''}

   \maketitle

\section{Introduction}

Since 1995, more than 120 planetary candidates have been 
discovered, mainly by radial velocity surveys. 
Although they provide a lot of information concerning 
the orbital parameters and the host star properties, such surveys do not yield 
 the real mass of the planet (only $m\sin i$) and do not give 
any information about its size. 
The discovery of HD209458 by Doppler measurements (Mazeh et al. \cite{mazeh}) 
and photometric transit (Charbonneau et al. \cite{charbonneau}; Henry et al. \cite{henry}) 
led to the first complete characterization of an extra-solar planet, 
illustrating the real complementarity of the two methods. 
The OGLE survey (Optical Gravitational Lensing Experiment) announced the 
detection of 137 short-period multi-transiting objects 
(Udalski et al., \cite{udalski1}, \cite{udalski2}, \cite{udalski3}, \cite{udalski4}).
Recently Konacki et al. (\cite{konacki1}, \cite{konacki2}), thanks to a Doppler 
follow-up, announced the characterization of the first extra-solar planet 
with the unexpected extremely short period of 1.2 day, much below the lower end of the period
distribution of planets detected by Doppler surveys (Udry et al. \cite{udry}). 
We present in this letter two new cases of extra-solar planets with very short orbital 
periods: OGLE-TR-113 and OGLE-TR-132.

\section{Observations and reductions}

The FLAMES facilities on the VLT (available since march 2003, Pasquini et al. 2002) is a 
very efficient way to conduct the Doppler follow-up of OGLE candidates. 
FLAMES is a multi-fiber link which feeds into the spectrograph UVES up to 
7 targets on a field-of-view of 25 arcmin diameter in addition to the 
simultaneous thorium calibration. The fiber link produces a stable 
illumination at the entrance of the spectrograph and permits the use of simultaneous 
ThAr calibration in order to track instrumental drift. Forty-five minutes on a 
$17^{th}$ magnitude star yield a signal-to-noise ratio (SNR) of about 8, corresponding 
to a photon noise uncertainty of about 30 {\ms} on a non-rotating K dwarf star.
We have obtained 3.2 nights in visitor mode in March 2004 on this instrument (program 72.C-0191) 
in order to observe  all OGLE candidates of the Carina field 
suspected to have a planetary companion.  
We present here the results and the analysis of three of these candidates, OGLE-TR-113, 
OGLE-TR-131 and OGLE-TR-132. 

The spectra obtained from the FLAMES+UVES spectrograph were extracted using 
the standard ESO-pipeline with bias, flat-field and background correction. Wavelength 
calibration was performed with ThAr spectra. The radial velocities were obtained by 
cross-correlation with a G2 digital mask. The instrumental drift was computed by cross-correlation 
of the simultaneous ThAr spectrum with a Thorium mask. Radial velocity errors were 
computed as a function of the SNR of the spectrum and the width ($FWHM$) of the 
Cross-Correlation Function (CCF) through the following relation based 
on photon noise simulations: $\sigma_{RV}=\,0.025\,FWHM / \,SNR$.

However, our measurement are clearly not photon noise limited and 
we added quadratically an error of 35 {\ms} in order to take into the account 
systematic errors probably due to wavelength calibration errors, fiber-to-fiber 
contamination, and residual cosmic rays. This value is based on our experience with FLAMES+UVES 
and are confirmed by the velocity residuals for OGLE-TR-131.

\section{Results}

\subsection{Radial velocities}

\begin{figure}
\resizebox{8.5cm}{!}{\includegraphics{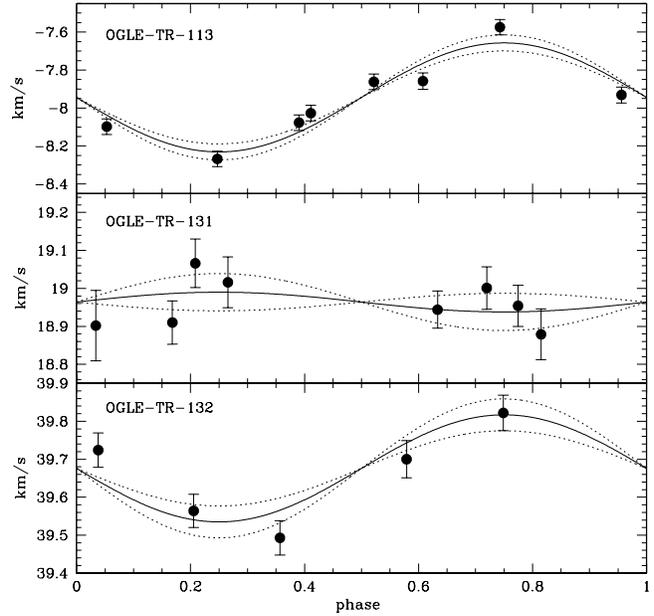}}
\caption{Phase-folded Doppler measurements of OGLE-TR-113, 131 and 132. 
The dotted lines correspond to fit curves for lower and upper 1-sigma intervals in semi-amplitude $K$.}
\label{doppler}
\end{figure}

Our radial velocity measurements are listed in Table \ref{tablevr}.
Figure \ref{doppler} shows 
the radial velocity data phased with the period and transit epoch 
from Udalski et al. (\cite{udalski3}, \cite{udalski4}). If the radial velocity variations are 
caused by the transiting objects, then phase $\phi=0$ must correspond to the passage 
of the curve at center-of-mass velocity with decreasing velocity, which provides a 
further constraint. We fitted the data with a sinusoid (assuming a circular orbit) 
and determined the velocity semi-amplitude $K$ and the center-of-mass velocity $V_0$. 
The orbital parameters are reported in Table~\ref{tablespectro}.

\begin{table}
\begin{tabular}{c c c c c c}
BJD & RV & depth & FWHM & SNR & $\sigma_{RV}$  \\ 
{[$-$2453000]} & [{\kms}] & [\%] & [{\kms}] & & [{\kms}] \\ \hline \hline
OGLE-113 & &  & & & \\ \hline
78.60419 & $-$7.862 & 40.23 & 9.4 & 11.0 & 0.041 \\
79.64235 & $-$8.268 & 40.07 & 9.5 & 11.7 & 0.040 \\
80.65957 & $-$7.931 & 39.73 & 9.6 & 10.2 & 0.042 \\
81.59492 & $-$7.858 & 38.94 & 9.5 &  9.3 & 0.043 \\
82.71279 & $-$8.077 & 40.19 & 9.5 & 11.5 & 0.041 \\
83.66468 & $-$8.098 & 39.60 & 9.4 & 10.8 & 0.041 \\
84.65149 & $-$7.574 & 40.91 & 9.4 & 12.0 & 0.040 \\
85.60720 & $-$8.027 & 40.32 & 9.4 & 11.5 & 0.041 \\ \hline
OGLE-131& & & & & \\ \hline
78.57421  & 18.902 &  20.40 & 11.0  & 3.2  &  0.093 \\
79.69280  & 18.944 &  36.18 & 10.5  & 7.5  &  0.049 \\
80.69588  & 18.910 &  33.25 & 10.8  & 6.0  &  0.057 \\
81.72914  & 19.001 &  32.90 & 10.2  & 5.8  &  0.056 \\
82.64055  & 19.066 &  31.21 & 10.6  & 4.9  &  0.064 \\
83.70039  & 18.954 &  33.55 & 10.1  & 6.2  &  0.054 \\
84.61585  & 19.016 &  27.87 & 10.3  & 4.5  &  0.067 \\
85.64254  & 18.879 &  28.34 & 10.6  & 4.6  &  0.067 \\ \hline
OGLE-132& &  & & & \\ \hline
81.72913  & 39.724  & 30.90  & 10.0  &  9.0   &  0.045 \\
82.64054  & 39.700  & 29.41  & 10.3  &  7.6   &  0.049 \\
83.70038  & 39.564  & 31.09  & 10.0  &  9.6   &  0.044 \\
84.61585  & 39.822  & 30.56  & 10.3  &  8.2   &  0.047 \\
85.64254  & 39.493  & 30.66  & 10.1  &  8.7   &  0.045 \\ 
\end{tabular}
\caption{Radial velocity measurements (in the barycentric frame) and CCF parameters 
for OGLE-TR-113, 131 and 132.}
\label{tablevr}
\end{table}

The existence of an orbital signal for OGLE-TR-113 is clear. 
For OGLE-TR-132, the reduced $\chi^2$ of a constant velocity curve without orbital motion 
is 33.1 ($P(\chi^2)\sim 10^{-6}$). Even if unrecognized systematics caused our error bars 
to be underestimated by a factor 1.5, the reduced $\chi^2$ would still be 14.7 ($P(\chi^2)=0.5$\%). As a 
foolproof check of the detection confidence, we also applied a bootstrap procedure to the data.
Bootstrapping gives an estimation of the significance of a signal without any assumption on the size of the uncertainties
(see e. g. Press et al. 1992). This yields a positive value of 
K in 97\% of the cases. Therefore, even with the assumption of large unrecognized systematics, 
the detection of orbital motion for OGLE-TR-132 on the correct period and phase is robust.
   
\begin{table*}
\begin{tabular}{c c c r r c c c c}
Name & P$_{OGLE}$ & T0$_{OGLE}$ & K & V$_0$ & O-C & $T_{eff}$ & $\log g$ & [Fe/H] \\ 
 & [days] & $-$2452000 &  [{\kms}] & [{\kms}]& [{\ms}] & & & \\  \hline \hline
OGLE-TR-113 & 1.4325 & 324.36394  &    0.287$\pm$0.042  & $-$7.944$\pm$0.027 & 66 & 4752$\pm$130 K& 4.50$\pm$0.53 & 0.14$\pm$0.14 \\
OGLE-TR-131 & 1.8699 & 324.94513  & $-$0.026$\pm$0.049 & 18.964$\pm$0.023 & 56 & 5244$\pm$136 K& 3.30$\pm$0.73 & 0.11$\pm$0.20\\
OGLE-TR-132 & 1.68965 & 324.70067 &   0.141$\pm$0.042  & 39.676$\pm$0.032 & 53 & 6411$\pm$179 K& 4.86$\pm$0.14$^{a}$ & 0.43$\pm$0.18 \\
\end{tabular}
\caption{Orbital and stellar spectroscopic parameters.$^{a}$ The unrealistically small value $\sigma_{\log g}=0.14$ for 132 was replaced with 
$\sigma_{\log g}=0.5$.}
\label{tablespectro}
\end{table*}

\subsection{Spectral line bisectors and blend scenarios}

\begin{figure}
\resizebox{8.5cm}{!}{\includegraphics{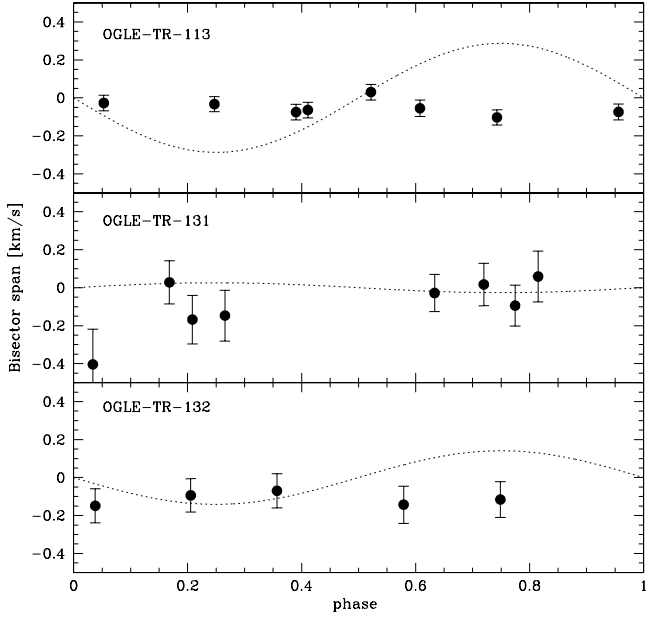}}
\caption{Bisector span [V$_{top}$ - V$_{bottom}$] of the three targets. Errors correspond to twice 
the radial velocity errors. Dotted lines correspond to the velocity signal from Fig. \ref{doppler}.}
\label{bisspan}
\end{figure}

It is known that in certain circumstances, the combination of a single star with a 
background unresolved eclipsing binary can mimic both a planet transit signal and 
velocity variations (Santos et al. \cite{santos2}). If the velocity variation are caused by 
a background binary system, however, the CCF bisector is expected to vary. In order to 
examine the possibility that the radial velocity variation 
is due to a blend scenario, we computed the CCF bisectors as described by 
Santos et al. (\cite{santos2}). Figure \ref{bisspan} indicates that 
there is no correlation of the line asymmetries with phase. 
Furthermore the CCF was computed with different 
masks without significant change in the radial velocity value (as discussed by Santos et al. 2002,
most blend scenarios produce mask-dependent velocities).
Moreover, a background binary of such short period would be expected to be synchronized, 
and thus to show a cross-correlation function very broadened by rotation.
Simulations show that a broad background CCF contaminating a narrow foreground
CCF is very inefficient in causing an apparent velocity variation. 
In order to provoke 
variations of the observed amplitude, any broadened background CCF would have to be
large enough to be clearly visible in the total CCF, which is not the case. Therefore, 
the scenario "foreground single star plus background short-period eclipsing binary" 
can be eliminated with confidence. While other more intricate scenarios could conceivably 
be possible, we were not able to contrive any that could explain both the photometric 
and the velocity signals while remaining credible.

\subsection{Spectral classification}

On the summed spectra, the intensity and equivalent width of some spectral lines were analyzed 
to give temperature, gravity and metallicity estimates for the primaries in the manner described in 
Santos et al. (\cite{santos3}). The results are given in Table~\ref{tablespectro}.

\subsection{light curve analysis and physical parameters}

The shape of the transit light curve depends in a non-linear way on 
 the latitude of the transit, the radius ratio $r/R$, the sum of masses $m+M$ and the primary radius $R$, 
where $R$, $M$, $r$, $m$ are the radius and mass values for the eclipsed and 
eclipsing bodies respectively. 
Synthetic transit curves computed with the procedure of Mandel \& Agol (\cite{mandel}) 
were fitted to the photometry data by least-squares. A quadratic limb 
darkening with $u1=u2=0.3$ was assumed (based on the values for HD209458 
used by Brown et al. \cite{brown}). Some possible sources of systematic 
uncertainties are not taken into account in this letter:
 variations in the limb darkening coefficients, 
possible contamination by background stars, uncertainties in the stellar 
evolution predictions, uncertainties in the orbital period. 
They will be included in the paper presenting the 
complete spectroscopic follow-up (Pont et al., in preparation).

The constraints on M and R were combined using a Bayesian procedure similar 
to that described in Pont \& Eyer (\cite{pont}): the posterior probability 
distribution of $M$ and $R$ was computed as the product of the combined 
likelihood from the light curve fit and the spectroscopic determinations of $T_{eff}$, $\log g$ and [Fe/H], and 
a prior probability distribution obtained from the Padua stellar evolution models (Girardi et al. 2002). 
The value of $r$ was then derived from $r/R$ and $R$, and $m$ from $M$ and 
the semi-amplitude of the radial velocity orbit, assuming circular Keplerian 
motion with $\sin i=1$ (The shortness of the period ensures that the orbits 
are circularized, and the presence of a transit indicates that $\sin i$ is 
very near to unity). Table \ref{tablezoo} summarize the resulting physical parameters 
of OGLE-TR-113 and OGLE-TR132 and their planetary companions.

\begin{figure}
\resizebox{8.5cm}{!}{\includegraphics[angle=-90]{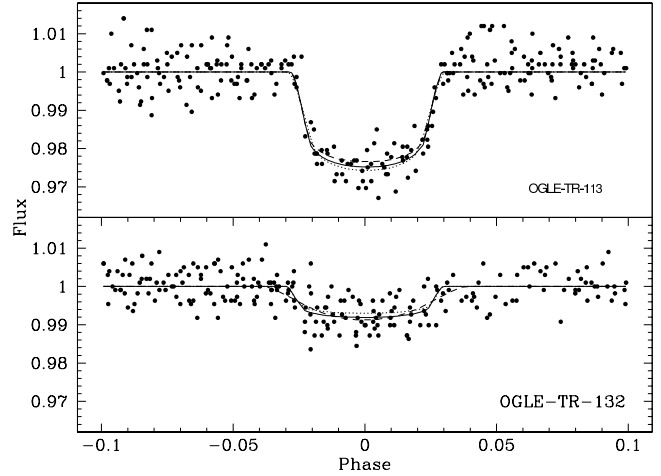}}
\caption{Phase-folded normalized light curve and best-fit transit curve (solid line) for OGLE-TR-113 
and OGLE-TR-132. The dotted and dashed lines correspond to fit curves for lower and upper 
1-sigma intervals in $r/R$.}
\label{ogle113}
\end{figure}

\section{Discussion and Conclusion}

\begin{table}
\begin{tabular}{c c c c c}
 Name& M & R & m & r \\ 
 & [$M_\odot$] & [$R_\odot$]& [$M_{Jup}$]& [$R_{Jup}$] \\ \hline \hline \\[-3mm]
OGLE-113  & 0.77$\pm$0.06 & 0.765$\pm$0.025        & 1.35$\pm$0.22 & 1.08$^{+0.07}_{-0.05}$ \\[1mm]
OGLE-132  & 1.34$\pm$0.10 & 1.41$^{+0.49}_{-0.10}$ & 1.01$\pm$0.31 & 1.15$^{+0.80}_{-0.13}$ \\[1mm]
\end{tabular}
\caption{Summary table with the physical parameters $M$, $R$, $m$ and $r$.}
\label{tablezoo}
\end{table}

The parameters of OGLE-TR-113b are very accurately defined, first because the 
transit shape is clearly delineated by the photometric data, second because there is only 
a narrow range of possible parameters allowed by stellar evolution models for 
a cool K dwarf. As a result, $r$ could be computed with a very 
small formal uncertainty.

No radial velocity variations were detected in OGLE-TR-131. Moreover, the 
spectroscopy indicates that it is most probably a sub-giant ($\log g=3.30\pm 0.73$), which renders 
the existence of a close companion unlikely.
OGLE-TR-131 is included here to show that our estimates of the radial velocity uncertainties  are
coherent with the residuals in the absence of detectable orbital motion.

\begin{figure}
\resizebox{8.5cm}{!}{\includegraphics{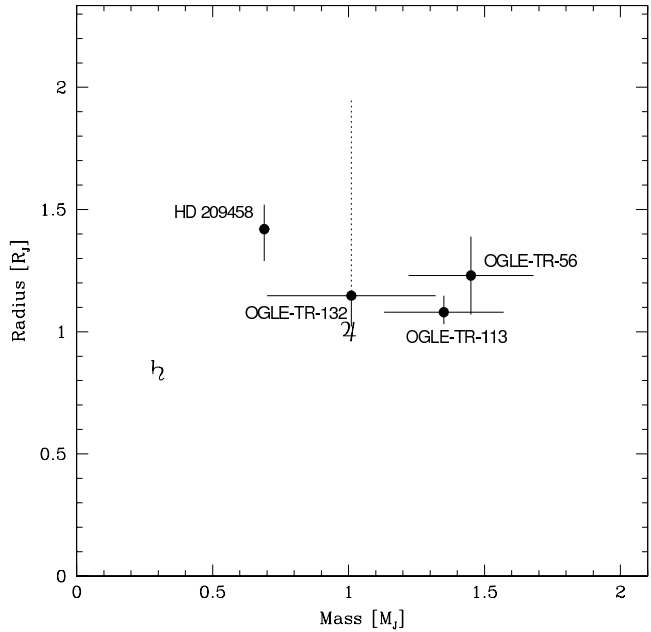}}
\caption{The four known transiting extrasolar planets plotted in the Mass-Radius diagram. The symbols indicate Jupiter and Saturn.}
\label{massradius}
\end{figure}

The photometric transit signal of OGLE-TR-132 is 
near the detectability limit, but 
the existence of a radial velocity variation at the precise period and phase 
of the transit gives confidence in the reality of the transiting companion.
In contrast with OGLE-TR-113, the parameters are not very well constrained. The transit 
shape is too poorly defined to constrain the transit latitude, so that there 
is a strong degeneracy between the impact parameter $b$ and the primary radius 
$R$. Furthermore the values of temperature and gravity measured from the 
spectra are compatible with a wide variety of young to evolved F dwarfs, with 
radii ranging from 1.3 to 1.9 $R_\odot$. As a result, the upper uncertainty 
interval on $r$ is wide, as indicated by the dotted error line in Fig. \ref{massradius}.

The semi-major axis of OGLE-TR-113b and OGLE-TR-132b are $a=0.0228 \pm 0.0006$ AU and 
$a=0.0306 \pm 0.0008$ AU respectively.


The detection of OGLE-TR-113b and OGLE-TR-132b show that the case of 
OGLE-TR-56 is not isolated and that "very hot Jupiters" (i.e. Jovian 
exoplanets with periods much smaller than 3 days) are not extremely 
uncommon. Therefore, the accumulation of hot Jupiters near periods of 3 
days (Udry et al. \cite{udry}) does not reflect an absolute limit for the existence of planets. 
The parent stars of OGLE-TR-56, 113 and 132 are very different, ranging from F 
to K dwarfs, indicating that very hot Jupiters are possible around different 
type of stars.

Fig. \ref{massradius} gives the mass-radius relation for the four known transiting exoplanets. 
It is noteworthy that the three OGLE objects do not seem to have such an inflated 
radius as HD209458b, despite their much shorter periods. 

The two new detections bring to 3 the number of "very hot  Jupiters" 
detected in the OGLE transit survey, among 155'000 light curves examined for 
transits. Given the geometric probability of transit ($\sim $17\% for $P=1.5$ days), this 
results implies a total number of 21 $\pm$ 10 very hot Jupiters among the targets. 
The detection completeness of the OGLE 
survey toward transiting very hot Jupiter should be computed with detailed 
simulations, but it may be quite high because for such low values of period 
the phase coverage is very good. Assuming a detection probability of 50\%, 
then the total number is $42 \pm 20$, therefore one in every 2500 to 7000 targets. Even if there 
is a proportion of giants stars among the OGLE candidate that would have 
been weeded out of the radial velocity surveys, the absence of "very hot 
Jupiters" among the $\sim$3000 field dwarfs surveyed in radial velocity in the 
solar neighborhood is therefore not incompatible with our result.
 This estimate also indicates that "very hot Jupiters" are not out of reach 
of future radial velocity surveys. It is noteworthy though that for such low periods, 
photometric transit surveys are a more efficient detection method than radial velocity monitoring.

\begin{acknowledgements}
We are grateful to C. Moutou for very useful comments and J. Smoker for support at Paranal. 
The data presented herein were obtained as part of an ESO visitor mode run (program 72.C-0191). 
F.P. gratefully acknowledges the support of CNRS through the fellowship program of CNRS. 
Support from Funda\c{c}\~ao para a Ci\^encia e Tecnologia (Portugal) to N.C.S. in the form 
of a scholarship is gratefully acknowledged.  

\end{acknowledgements}


\begin{thebibliography}{}

\bibitem[2001]{brown} Brown, T.M., Charbonneau, D., Gilliland, R., et al., 2001, ApJ, 552, 699 

\bibitem[2000]{charbonneau} Charbonneau, D., Brown, T.M., Latham, D., \& Mayor, M., 2000, ApJ, 529, L45

\bibitem[2002]{girardi} Girardi, M., Manzato, P., Mezzetti, M., et al., 2002, ApJ, 569, 720

\bibitem[2000]{henry} Henry, G.W., Marcy, G.W., Butler, R.P. \& Vogt, S.S., 2000, ApJ, 529, L41

\bibitem[2003]{konacki1} Konacki, M., Torres, G., Jha, S., et al., 2003a, Nature, 421, 507

\bibitem[2004]{konacki2} Konacki, M., Torres, G., Sasselov, D., et al., 2004, ApJ, in press

\bibitem[2002]{mandel} Mandel, K. \& Agol, E., 2002, ApJ, 580, 171

\bibitem[2000]{mazeh} Mazeh, T., Naef, D., Torres, G., et al., 2000, ApJ, 532, L55

\bibitem[2002]{pasquini} Pasquini, F., Avila, G., Blecha, A., et al., 2002, The Messenger, 110, 1

\bibitem[2004]{pont} Pont, F. \& Eyer, L., 2004, MNRAS, in press (astrop-ph/0401418)

\bibitem[1992]{press} Press, W.H., et al., 1992, {\it Numerical recipes}, Cambridge University Press, p. 690

\bibitem[2002]{santos2} Santos, N.C., Mayor, M., Naef, D., et al., 2002, A\&A, 392, 215

\bibitem[2004]{santos3} Santos, N.C., Israelian, G., Mayor, M., 2004, A\&A, 415, 1153

\bibitem[2003]{sirko} Sirko, E. \& Paczynski, B., 2003, ApJ, 592, 1217

\bibitem[2002a]{udalski1} Udalski, A., Paczynski, B., Zebrun, K.,et al., 2002a, Acta Astron., 52, 1

\bibitem[2002b]{udalski2} Udalski, A., Zebrun, K., et al., 2002b, Acta Astron., 52, 115

\bibitem[2002c]{udalski3} Udalski, A., Zebrun, K., et al., 2002c, Acta Astron., 52, 317

\bibitem[2003]{udalski4} Udalski, A., Pietrzynski, G, et al., 2003, Acta Astron., 53, 133

\bibitem[2003]{udry} Udry, S., Mayor, M., Santos, N.C., 2003, A\&A, 407, 369




\end{thebibliography}
\end{document}